\documentclass[prl,aps,epsfig,showpacs]{revtex4} 

\usepackage{amsfonts}
\usepackage[dvips]{graphicx}
\usepackage{mathrsfs}
\usepackage[intlimits]{amsmath}
\usepackage[colorlinks, citecolor=red]{hyperref}
\usepackage{graphicx}  
\usepackage{dcolumn}   
\usepackage{bm}        
\usepackage{amssymb}   
\usepackage{mathtools}
\usepackage{braket}

\usepackage[english]{babel}
\usepackage{blindtext}

\makeatletter
\def\blfootnote{\xdef\@thefnmark{}\@footnotetext}
\makeatother

\newcommand{\overbar}[1]{\mkern 2.3mu\overline{\mkern-2.3mu#1\mkern-2.3mu}\mkern 2.3mu}

\begin{document}

\title{Quantum Discriminant Analysis for Dimensionality Reduction and
Classification}
\author{Iris Cong$^{1,3}$ and Luming Duan$^{2,3}$}
\affiliation{$^{1}$Department of Computer Science, University of California, Los Angeles,
CA 90095}
\affiliation{$^{2}$Department of Physics, University of Michigan, Ann Arbor, Michigan
48109, USA}
\affiliation{$^{3}$Center for Quantum Information, IIIS, Tsinghua University, Beijing
100084, PR China}

\begin{abstract}
We present quantum algorithms to efficiently perform discriminant analysis
for dimensionality reduction and classification over an exponentially large
input data set. Compared with the best-known classical algorithms, the
quantum algorithms show an exponential speedup in both the number of
training vectors $M$ and the feature space dimension $N$. We generalize the
previous quantum algorithm for solving systems of linear equations [\textit{%
Phys. Rev. Lett.} \textbf{103}, 150502 (2009)] to efficiently implement a
Hermitian chain product of $k$ trace-normalized $N\times N$ Hermitian
positive-semidefinite matrices with time complexity of $O(\log (N))$. Using
this result, we perform linear as well as nonlinear Fisher discriminant
analysis for dimensionality reduction over $M$ vectors, each in an $N$%
-dimensional feature space, in time $O(p \text{ polylog} (MN)/\epsilon ^{3})$, where $%
\epsilon $ denotes the tolerance error, and $p$ is the number of principal projection directions desired. We also present a quantum
discriminant analysis algorithm for data classification with time complexity
$O(\log (MN)/\epsilon ^{3})$.
\end{abstract}

\maketitle

\section{I.  Introduction}
{ \setcitestyle{super}
\blfootnote{* The authors can be reached at: iriscong23$@$ucla.edu, lmduan$@$umich.edu}
}

With the rise in the fields of big data analysis and machine learning in the
modern era, techniques such as dimensionality reduction and classification
have gained significant importance in the information sciences. In machine
learning and statistical analysis problems, when input vectors are given in
an extremely large feature space, it is often necessary to reduce the data
to a more manageable dimension/size before manipulation or classification.
One classical example is in the problem of face recognition \cite{Belhumeur,
Cheng}, where the size of the feature space is determined by a huge number
of pixels representing each face. More recent applications are also seen in
fields of medical imaging. For instance, \cite{Uetani} shows the necessity
for dimensionality reduction in diagnosing cases of liver cirrhosis. Also,
\cite{Dai} shows the importance of dimensionality reduction for early
Alzheimer's disease detection.

One widely used technique for dimensionality reduction is principal
components analysis (PCA), where the data is projected onto the directions
of maximal variance. However, a significant disadvantage of PCA is that it
looks only at the overall data variance, and does not consider the class
data. The extreme example of this would occur if the overall data variance is in exactly the same direction as the maximal within-class data variance, but orthogonal to the direction of maximal between-class data variance. In such a case, it is possible for a PCA projection to completely overlap the
data from different classes, making it impossible to use the projected data
to perform future discriminations. Fisher's linear discriminant analysis
(LDA) is a technique developed to overcome this problem by instead
projecting the data onto directions that maximize the between-class
variance, while minimizing the within-class variance of the training data.
It is hence not surprising that LDA is shown to be more effective than PCA
in machine learning problems involving dimensionality reduction before
classification \cite{Belhumeur, Cheng}.

Another common application of discriminant analysis is to use it as a
classifier itself, where labeled training vectors are presented as input and
new cases must be efficiently assigned to their respective classes. The
discriminant analysis classifier has recently been used in medical analysis,
such as in analyzing electromyography (EMG) signals \cite{Naik, Krusienski},
lung cancer classification \cite{Suarez-Cuenca}, and breast cancer diagnosis
\cite{Esener}. While other algorithms such as the support vector machine
(SVM) reach similar accuracy rates as the discriminant analysis classifier
\cite{Alkan,Dixon}, studies \cite{Dixon} show that discriminant analysis is
a significantly more stable model. This is because the separating hyperplane
chosen by the SVM can depend only on a few support vectors, subjecting it to
high variance in the case of training vector errors. On the other hand,
since discriminant analysis bases its classification on the entire class
means and variances, it tends to show less fluctuation and is potentially
more robust in the presence of error.

A significant drawback of discriminant analysis in both dimensionality
reduction and classification is the time complexity. Even the best existing
classical algorithms for LDA dimensionality reduction require time $O(Ms)$
\cite{Cai}, where $M$ is the number of training vectors given, and $s$ is
the sparseness (maximum number of nonzero components) of each feature
vector. For large data sets in high dimensions, this can scale rather
poorly, especially since it is often hard to guarantee the sparseness of
training vectors. While quantum algorithms have been designed to
exponentially speed up PCA to be polylogarithmic in the number of input
vectors and their dimension \cite{Lloyd14}, no such work has been done yet
to speed up LDA. In Section III, we provide a quantum algorithm for LDA
polylogarithmic in both $M$, the number of training vectors, and $N$, the
initial feature space dimension, regardless of the sparseness of the
training vectors.

Similarly, while a quantum algorithm has been presented to provide
exponential speedup in SVM classification \cite{LloydSVM}, no work has been
done yet for the discriminant analysis classifier. The best known algorithms
for the classical discriminant analysis classifier also require time
polynomial in $M$ and $N$ (see Section II below), which again can scale
rather poorly. In Section IV, we provide a quantum algorithm for discriminant
analysis classification logarithmic in both the number of input vectors $M$
and the dimension $N$.

This paper is arranged as follows: In Sec. II we briefly review the
classical discriminant analysis algorithms for dimensionality reduction and
data classification. In Sec. III and IV, we present our major results of
quantum discriminant analysis algorithms for dimensionality reduction as
well as classification. The detailed proof of Theorem 1 in Sec. III, which is
about efficient quantum implementation of a Hermitian chain product of $k$
trace-normalized $N\times N$ Hermitian positive-semidefinite matrices, is included
in the appendix.

\section{II.  Review of Classical Discriminant Analysis}

\subsection{Dimensionality Reduction}

The classical LDA dimensionality reduction algorithm is designed to return
the directions of projection that maximize the between-class variance (for
class discrimination), but minimize the within-class variance. With this result, in big data problems as listed in Section I, the vectors in a high-dimensional feature space may be projected onto a lower-dimensional subspace (spanned by the returned optimal unit vectors), so that less resources may be used to store the same amount of information. Suppose we are given $M$ (real-valued) input data vectors $\{x_i \in \mathbb{R}^N: 1 \leq i \leq M\}$ each belonging to one of $k$ classes. Let $\mu_c$ denote the within-class mean (centroid) of class $c$%
, and $\mkern 2.3mu\overline{\mkern-2.3mu x\mkern-2.3mu}\mkern 2.3mu$
denotes the mean/centroid of all data points $x$. Furthermore, let 

\begin{equation}
S_B = \sum_{c = 1}^{k} (\mu_c - \mkern 2.3mu\overline{\mkern-2.3mu x\mkern%
-2.3mu}\mkern 2.3mu)(\mu_c - \mkern 2.3mu\overline{\mkern-2.3mu x\mkern-2.3mu%
}\mkern 2.3mu)^T
\end{equation}

\noindent
denote the between-class scatter matrix of the dataset, and let

\begin{equation}
S_W = \sum_{c = 1}^{k} \sum_{x \in c} (x - \mu_c)(x - \mu_c)^T.
\end{equation}

\noindent
denote the within-class scatter matrix. The goal is then to find a direction of projection $w \in \mathbb{R}^N$ that maximizes the between-class variance $w^T S_B w$ relative to the within-class variance $w^T S_W w$. Mathematically, assuming that the classes have approximately multivariate
Gaussian distribution with similar covariances, this is the problem of
maximizing the objective function (commonly known as \textit{Fisher's
discriminant})

\begin{equation}
J(w) = \frac{w^T S_B w}{w^T S_W w}.
\end{equation}

Since the expression for $J(w)$ is invariant under constant rescaling of $w$%
, it is clear that the maximization problem given in (3) is equivalent to
the optimization problem

\begin{equation}
\min_{w} -w^T S_B w
\end{equation}

\begin{equation}
\text{subject to } w^T S_W w = 1.
\end{equation}

\noindent We are thus minimizing the Lagrangian \cite{Welling}

\begin{equation}
\mathcal{L}_P = -w^T S_B w + \lambda(w^T S_W w - 1)
\end{equation}

\noindent where $\lambda$ is the desired Lagrange multiplier. By the Karush-Kuhn-Tucker conditions \cite{Boyd}, this means that

\begin{equation}
S_W^{-1}S_B w = \lambda w.
\end{equation}

\noindent It follows that $w$ is an eigenvector of $S_W^{-1}S_B$. By
plugging (7) back into the objective function $J(w)$, we get $J(w) = \lambda$%
. Hence, we choose $w$ to be the principal eigenvector.

The above procedure generalizes easily to higher-dimensional projection subspaces. In this case, we seek $p$ vectors which form a basis for our projection subspace; this corresponds to maximizing the discriminant

\[
J(W) = \frac{W^T S_B W}{W^T S_W W}
\]

\noindent
where $W$ is the $N \times p$ matrix whose columns are the basis vectors. Using the same analysis as above, one can show that the columns of $W$ will be the eigenvectors corresponding to the $p$ largest eigenvalues of $S_W^{-1}S_B$, as in the case of PCA.

\subsection{Classification}

Although its most widely-used application is probably in dimensionality
reduction, discriminant analysis is also commonly used to directly perform
data classification. For classification, one constructs the \textit{%
discriminant functions} for each class $c$

\begin{equation}
\delta_c (x) = x^T \Sigma_c^{-1} \mu_c - \frac{1}{2} \mu_c^T \Sigma_c^{-1}
\mu_c + \log \pi_c
\end{equation}

\noindent where $\Sigma_c$ is the covariance matrix for class $c$, $\mu_c$
is the class mean for $c$ as before, and $\pi_c$ is the prior probability
for classifying into class $c$ \cite{Hastie}. Given a vector $x$, it is then
classified into the class $c = \text{argmax}_c \delta_c(x)$. From the
training vector data, if $M_c$ is the number of training vectors belonging
to class $c$, we can approximate $\pi_c = M_c/M$ for simplicity, i.e. the probability of classifying to a certain class $c$ is directly proportional to the fraction of training vectors belonging to $c$ \cite{Hastie}. Assuming
multivariate Gaussian distributions for each class, we also estimate

\begin{equation}
\Sigma_c = \frac{1}{M_c - 1} \sum_{x \in c} (x - \mu_c) (x - \mu_c)^T.
\end{equation}

\noindent Note that in the special case where the covariance matrices are
all approximately equal (i.e. $\Sigma _{c}\approx \Sigma $ $\forall c$), $%
\Sigma $ is proportional to the scatter matrix $S_{W}$ given by Eq. (2). In
this special case, the functions $\delta _{c}$ are known as \textit{linear
discriminant functions}. In our paper, we present a quantum algorithm to
solve the more general case, known as \textit{quadratic} discriminant
analysis (QDA), in time logarithmic in both the number of input vectors $M$
and their dimension $N$. Our algorithm will be easily applicable to the
special case of LDA classification. This provides exponential speedup over
the fastest existing algorithms, since the classical construction/inversion
of $\Sigma _{c}$ to evaluate the discriminant functions must require time
polynomial in both $M$ and $N$.

\section{III.  Quantum LDA Algorithm: Dimensionality Reduction}

\subsection{Assumptions and Initialization}

The quantum Principal Components Analysis algorithm of \cite{Lloyd14} presents methods for processing input vector data if the covariance matrix of the data is efficiently obtainable as a density matrix, under specific assumptions about the vectors given in quantum mechanical form. While our major contributions are also in the processing mechanisms of the within- and between-class covariance matrices, we will describe how to obtain this density matrix under certain assumptions about the input data, like in Refs. \cite{Lloyd14,HHL,Lloyd13}.

In our algorithm, similar to the assumptions made in Refs. \cite{Lloyd14,HHL,Lloyd13}, we will assume we have quantum access to the training vector data
in a quantum RAM (as described in \cite{Giovannetti08}). We will assume that each training vector is stored in the quantum RAM in terms of its difference from the class means. That is, if a training vector $x_j$ belongs to class $c_j$ with centroid $\mu_{c_j}$, we have the Euclidean norm and complex-valued components of the difference vector $d_j = x_j - \mu_{c_j}$ stored as floating-point numbers in quantum RAM in polar form (alternatively, if the input is presented directly as the training vectors $x_j$ and the class means $\mu_c$, we may first perform a component-wise subtraction of the given floating-point numbers, by \cite{Nakahara}). Following the methodology of \cite{Lloyd13,Lloyd14,LloydSVM}, we will assume the following oracle:

\begin{equation*}
\mathcal{O}_1 (\ket{j}\ket{0}\ket{0}\ket{0}) \rightarrow \ket{j}\ket{\| x_j - \mu_{c_j}\| }\ket{x_j - \mu_{c_j}}\ket{c_j}
\end{equation*}

\noindent to get the $j^{\text{th}}$ training vector and its class $c_{j}$, where $\ket{x_j - \mu_{c_j}}$ has already been normalized to one.
Similarly, we also assume that we are given the quantum representations of the class
centroids $\ket{\mu_c}$, in terms of their differences from the overall training vector mean $%
\ket{\overbar{x}}$. That is, if $D_c = \mu_c - \overbar{x}$, we assume the oracle

\begin{equation*}
\mathcal{O}_2 (\ket{c} \ket{0}\ket{0}) \rightarrow \ket{c} \ket{\|\mu_c - \overbar{x} \|} \ket{\mu_c - \overbar{x}}
\end{equation*}

\noindent
where we similarly assume that $\ket{\mu_c - \overbar{x}}$ has been normalized to one. These oracles could, as an example, be realizable if the input data is presented in this form as the output of a preceding quantum system, or if the vector components are presented as floating-point numbers in the quantum RAM, and the sub-norms of the vectors can be estimated efficiently \cite{Lloyd13,Grover,Kaye,Soklakov}. The oracles $\mathcal{O}_{1}$ and $\mathcal{O}_{2}$ allow
us to construct density matrices proportional to $S_{B}$ and $S_{W}$ as
follows: Let

\begin{equation}
\ket{\Psi_1} = \mathcal{O}_2 \left(\frac{1}{\sqrt{k}} \sum_{c=1}^{k} \ket{c}%
\ket{0} \ket{0}\right) = \frac{1}{\sqrt{k}} \sum_{c=1}^{k} \ket{c} %
\ket{\|\mu_c - \overbar{x} \|} \ket{\mu_c - \overbar{x}}
\end{equation}

\begin{equation}
\ket{\Phi_1} = \mathcal{O}_1 \left(\frac{1}{\sqrt {M}}\sum_{j=1}^{M} \ket{j} %
\ket{0} \ket{0}\ket{0}  \right) = \frac{1}{\sqrt M} \sum_{j=1}^{M} \ket{j}%
\ket{\| x_j - \mu_{c_j}\| }\ket{x_j - \mu_{c_j}} 
\end{equation}

\noindent
By \cite{Lloyd13,Grover,Kaye,Soklakov}, if the norms of the vectors form an efficiently integrable distribution, we can obtain the states

\begin{equation}
\ket{\Psi_2} = \frac{1}{\sqrt{A}}  \sum_{c=1}^{k} \|\mu_c - \overbar{x} \| \ket{c} %
\ket{\|\mu_c - \overbar{x} \|} \ket{\mu_c - \overbar{x}}
\end{equation}

\begin{equation}
\ket{\Phi_2} = \frac{1}{\sqrt B} \sum_{j=1}^{M} \| x_j - \mu_{c_j}\| \ket{j}%
\ket{\| x_j - \mu_{c_j}\| }\ket{x_j - \mu_{c_j}}  \ket{c_j}
\end{equation}

\noindent
where $A = \sum_{c=1}^{k} \|\mu_c - \overbar{x} \|^2$, $B = \sum_{j=1}^{M} \| x_j - \mu_{c_j}\|^2$. 

In both cases, we now take the partial trace over the first register. Then, for the
state of Eq. (12), the density matrix of the second register \cite{Lloyd14} is
given by

\begin{equation}
S_B = \frac{1}{A} \sum_{c=1}^{k} \| \mu_c - \overbar{x} \|^2 \ket{\mu_c - \overbar{x}}\bra{\mu_c -
\overbar{x} }
\end{equation}

\noindent and for the state of Eq. (13), the density matrix of the second register is
given by

\begin{equation}
S_{W}= \frac{1}{B}\sum_{c=1}^{k}\sum_{i\in c} \| x_i - \mu_c \|^2 \ket{x_i - \mu_c}\bra{x_i -
\mu_c}.
\end{equation}

\noindent Assuming our oracles $\mathcal{O}_1$ and $\mathcal{O}_2$, we can hence construct the Hermitian operators $S_B$, $S_W$
in time $O(\log (MN))$.

\subsection{LDA Approach}

Having initialized the means and operators $S_{B}$, $S_{W}$, our main task
will be to solve the eigenvector problem of (7). This problem would be
simpler if only $S_{W}^{-1}S_{B}$ were Hermitian positive semidefinite.
However, we can reduce this problem to an eigenvalue problem for a Hermitian
density matrix: specifically, since $S_{B}$ is Hermitian positive semidefinite,
letting $w=S_{B}^{-1/2}v$ reduces (7) to the eigenvalue problem \cite%
{Welling}

\begin{equation}
S_B^{1/2}S_W^{-1}S_B^{1/2} v = \lambda v
\end{equation}

\noindent To apply the quantum phase estimation algorithm to solve (16), we
must first be able to construct the density matrix $%
S_B^{1/2}S_W^{-1}S_B^{1/2}$. In the following section, we present a more
general theorem that can be applied to construct this density matrix.

\subsection{Implementing the Hermitian chain product}

In this section, we state a theorem to construct the density matrix corresponding to the Hermitian chain product

\begin{equation}
[f_k(A_k)...f_1(A_1)][f_k(A_k)...f_1(A_1)]^\dagger
\end{equation}

\noindent to error $\epsilon$, for arbitrary normalized $N \times N$
Hermitian positive semidefinite matrices $A_1, ... A_k$, and functions $f_1,
... f_k$ with convergent Taylor series. The derivation of this theorem
follows the method presented in \cite{HHL}, and is presented in Appendix A.

\vspace{4mm} \noindent \textbf{Theorem 1:} \textit{Let $A_{1},...A_{k}$ be $k
$ normalized Hermitian positive semidefinite matrices whose quantum forms
can be constructed in time $O(\log (N))$ (e.g., by visits to a quantum RAM)
and let $f_{1},...f_{k}$ be $k$ functions with convergent Taylor series. Let
$\{\lambda _{jl}\}_{l=1}^{N}$ denote the eigenvalues of matrix $A_{j}$. Then
the Hermitian operator in {\normalfont Eq. (17)} can be implemented in time}

\begin{equation}
O\left(\frac{\log(N)}{\epsilon^3} \cdot \sum_{j=1}^k \kappa_j^2 \cdot \left(
\frac{\max_l |f_1(\lambda_{1l})|}{\min_l |f_1(\lambda_{1l})|} \right)
\prod_{j=2}^k \left( \frac{\max_l |f_j(\lambda_{jl})|}{\min_l
|f_j(\lambda_{jl})|} \right)^2 \right).
\end{equation}

\noindent \textit{where $\kappa_j$ is the condition number for matrix $A_j$,
i.e. the ratio of the largest to smallest eigenvalue of $A_j$. More
generally, if construction of each matrix $A_j$ takes time $O(X)$, the
operator can be implemented in time}

\begin{equation}
O\left(\frac{X}{\epsilon^3} \cdot \sum_{j=1}^k \kappa_j^2 \cdot \left( \frac{%
\max_l |f_1(\lambda_{1l})|}{\min_l |f_1(\lambda_{1l})|} \right)
\prod_{j=2}^k \left( \frac{\max_l |f_j(\lambda_{jl})|}{\min_l
|f_j(\lambda_{jl})|} \right)^2 \right).
\end{equation}

We note that this provides exponential speedup over classical algorithms, as the optimal classical algorithm for multiplication of non-sparse $N \times N$ matrices requires time $O(N^{2.3737})$ \cite{Williams}.

\subsection{Finding the principal eigenvectors}

For LDA, we apply the theorem presented in the previous section to obtain
the matrix product $S_{B}^{1/2}S_{W}^{-1}S_{B}^{1/2}$. Specifically, we use $%
A_{1}=S_{W}$, $A_{2}=S_{B}$, $f_{1}(X)=X^{-1/2}$, and $f_{2}(X)=X^{1/2}$. To
avoid exponential complexity in the case of exponentially small eigenvalues,
we adopt a technique used in \cite{LloydSVM} by pre-defining an effective
condition number $\kappa _{\text{eff}}$ and taking into account only
eigenvalues in the range $[1/\kappa _{\text{eff}},1]$ for phase estimation.
(Typically, one may take $\kappa _{\text{eff}}=O(1/\epsilon )$, because $\kappa _{\text{eff}}$ is a limit to the amount of eigenvalues considered in phase estimation, which should be proportional to the error tolerance). By our initialization procedures, preparation of $S_{B}$ and $S_{W}$ take time $O(\log (MN))$, and by
definition of $f_{1}$, $f_{2}$, and $\kappa _{\text{eff}}$,

\begin{equation*}
\frac{\max_l |f_1(\lambda_{1l})|}{\min_l|f_1(\lambda_{1l})|} = \frac{\max_l
|f_2(\lambda_{2l})|}{\min_l|f_2(\lambda_{2l})|} = \kappa_{\text{eff}}^{1/2}.
\end{equation*}

\noindent Hence, by (19) we can obtain $S_{B}^{1/2}S_{W}^{-1}S_{B}^{1/2}$ in
time $O(\log (MN)\kappa _{\text{eff}}^{3.5}/\epsilon ^{3})$. Using the
matrix exponentiation technique presented in \cite{Lloyd14}, we can then
apply quantum phase estimation to obtain an approximation to the state

\begin{equation*}
\rho = \sum_i \lambda_i \ket{v_i} \bra{v_i} \otimes \ket{\lambda_i} \bra{\lambda_i}
\end{equation*}

\noindent
where $\lambda_i$ and $v_i$ are the eigenvalues and eigenvectors of $S_{B}^{1/2}S_{W}^{-1}S_{B}^{1/2}$. If the $p$ principal (largest) eigenvalues are polynomially small, sampling produces the corresponding $p$ principal eigenvectors $v_{r}$ of $S_{B}^{1/2}S_{W}^{-1}S_{B}^{1/2}$ in time $O(p\text{ polylog} (MN)/\epsilon ^{3})$
\cite{Cleve}. (If all eigenvalues are indeed super-polynomially small, there are typically no suitable directions for discriminant analysis: all directions would be essentially the same in preserving between-class vs. within-class data). Finally, having solved the eigenvalue problem of (16), we
again use the technique of the previous section to obtain the eigenvectors

\begin{equation}
w_r = S_B^{-1/2} v_r
\end{equation}

\noindent of $S_W^{-1} S_B$. After obtaining these principal eigenvectors,
the data can be projected onto the dimensions of maximal between-class
variance and minimal within-class variance. At this stage, one has
effectively performed dimensionality reduction, and can now easily
manipulate the data with existing tools, e.g. a classifier (see \cite%
{LloydSVM} or Section IV below).

\vspace{5.5mm}

\noindent\fbox{
\parbox[c][]{6.85in}{
\vspace{3.5mm}
\begingroup
\leftskip2.6mm
\rightskip\leftskip
{\bf \underline{Algorithm 1: Quantum LDA Dimensionality Reduction}}

\vspace{3mm}
\noindent
{\it Step 1}: Initialization. By querying the quantum RAM/oracles,  construct the Hermitian positive semidefinite operators $S_B$ and $S_W$ as given by Equations (14) and (15), in time $O(\log(MN))$.

\vspace{2mm}
\noindent
{\it Step 2}: Since $S_B$ and $S_W$ are Hermitian positive semidefinite, use the method of \cite{Lloyd14} to exponentiate these operators. Apply the generalized matrix chain algorithm of Theorem 1 to implement $S_B^{1/2}S_W^{-1}S_B^{1/2}$ in time $O(\log(MN) \kappa_{\text{eff}}^{3.5} / \epsilon^3)$.

\vspace{2mm}
\noindent
{\it Step 3}: Since $S_B^{1/2}S_W^{-1}S_B^{1/2}$ is Hermitian positive semidefinite, use the method of \cite{Lloyd14} to exponentiate this operator. Use quantum phase estimation methods and sample from the resulting probabilistic mixture to then obtain the $p$ principal eigenvalues/eigenvectors $v_r$, in time $O(p \text{ polylog}(MN)/\epsilon^3)$.

\vspace{2mm}
\noindent
{\it Step 4}: Apply $S_B^{-1/2}$ to the $v_r$'s (by the algorithm of Theorem 1) to get desired directions $w_r = S_B^{-1/2} v_r$, in time $O(p\log (MN) \kappa_{\text{eff}}^3 / \epsilon^3)$.

\vspace{2mm}
\noindent
{\it Step 5}: Project data onto the $w_r$'s for dimensionality reduction, or otherwise work in the directions of maximal class discrimination.

\endgroup

\vspace{5.5mm} 
} }

\vspace{1mm} 

\subsection{Algorithmic Complexity for Dimensionality Reduction}

Algorithm 1 above shows the pseudocode for our LDA algorithm. Step 1
(initialization) takes time $O(\log (MN))$ with our quantum oracles.
Implementing the operator $S_{B}^{1/2}S_{W}^{-1}S_{B}^{1/2}$ takes time $%
O(\log (MN)\kappa _{\text{eff}}^{3.5}/\epsilon ^{3})$, and finding its
principal eigenvectors then takes $O(\log (MN)/\epsilon ^{3})$. Finally,
Step 4 takes time $O(\log (MN)\kappa _{\text{eff}}^{3}/\epsilon ^{3})$ to
apply $S_{B}^{-1/2}$ to the $v_{r}$'s and obtain the eigenvectors of $%
S_{W}^{-1}S_{B}$. Hence, we can add these to get the total runtime:

\vspace{4mm} \noindent \textbf{Theorem 2:} \textit{The quantum LDA algorithm
presented in this paper (with pseudocode given by Algorithm 1) can be
implemented in time polylogarithmic in both the number of input vectors $M$ and
the input vector dimension $N$. Specifically, it has a runtime of}

\begin{equation}
O\left( p \text{ polylog} (MN) \kappa_{\text{eff}}^{3.5} / \epsilon^3 \right)
\end{equation}

\noindent \textit{where $\kappa_{\text{eff}}$ is a pre-defined condition
number restricting the range of eigenvalues considered for phase estimation,
typically taken to be $O(1/\epsilon)$, and $p$ is the number of principal eigenvectors.}

\vspace{3mm}
Note that the complexity presented here in (21) is
polylogarithmic in both the number of input vectors $M$ and their dimension $N$,
regardless of training vector sparseness.

\subsection{Nonlinear/Kernel Fisher Discriminant Analysis}

Certainly, in many real-world cases, a straightforward linear discriminant
may not be sufficient. Classically, a simple generalization is known as
kernel Fisher discriminant analysis (kernel FDA), where the input vectors
are first mapped (nonlinearly) by a function $\phi :x_{j}\rightarrow \phi
(x_{j})$ into a higher-dimensional feature space $\mathcal{F}$. The linear
discriminant corresponding to $J(w)$ in the feature space then becomes
nonlinear in the original space. In the classical case, if the dimension of $%
\mathcal{F}$ is too large, it becomes computationally infeasible to perform
operations such as matrix multiplication or exponentiation on the resulting
large covariance matrices $S_{B}^{\phi }$ and $S_{W}^{\phi }$. Instead, one
must find workarounds such as kernel methods, and perform reductions so that
the algorithm involves only dot products in the feature space \cite{Mika},
but this may seriously limit the potential choices of mapping $\phi $. In
the quantum case, however, we can directly perform the LDA analysis in the
higher-dimensional feature space. As long as the dimension of $\mathcal{F}$
scales polynomially with the original input dimension, our algorithmic
performance will be affected only by a constant factor. This allows for a
much wider range of mappings $\phi $ into the feature space.

\section{IV.  Quantum Discriminant Analysis Algorithm for Classification}

\subsection{Algorithm}

We now present an efficient quantum algorithm for the classification of an
exponentially large data set by quadratic discriminant analysis (QDA), and our results can easily be applied
to perform LDA classification. As before, we assume that the class means and training vector data are given with their norms and components stored as floating-point numbers in quantum RAM. We again assume the oracle $\mathcal{O}_2$ to obtain the class means of the training vectors $\mu_c$ and their norms, and in this section, we further assume that we can obtain
the $j^{\text{th}}$ training vector of each class. As in Section III initialization or Refs. \cite{Lloyd13,Lloyd14,LloydSVM}, our oracle gives the vectors $x_{c,j}$ are given as their difference from their class means (as in Section III, this may also be obtained from the stored floating-point numbers if necessary). Specifically, we assume the oracle

\begin{equation*}
\mathcal{O}_3(\ket{c}\ket{j}\ket{0}\ket{0}) \rightarrow \ket{c}\ket{j}\ket{\| x_{c,j} - \mu_c \|}\ket{x_{c,j} - \mu_c}%
.
\end{equation*}

\noindent
As in Section III, $\ket{x_{c,j} - \mu_c}$ has been normalized to one.
Finally, we now assume that for each class $c$, we are given the
number of training vectors $M_c$ belonging to that class.

For QDA, we use the oracles to construct for each class $c$ the Hermitian
positive semidefinite operator

\begin{equation}
\Sigma _{c}= \frac{1}{A_c} \sum_{j=1}^{M_{c}} \| x_{c,j} - \mu_c \|^2 \ket{x_{c,j} - \mu_c}%
\bra{x_{c,j} - \mu_c}.
\end{equation}

\noindent
where $A_c = \sum_{j=1}^{M_{c}} \| x_{c,j} - \mu_c \|^2$. To do this, we first call $\mathcal{O}_{3}$ on the state $\frac{1}{%
\sqrt{M_{c}}}\sum_{j=1}^{M_{c}}\ket{c}\ket{j}\ket{0}\ket{0}$ to obtain the register
$\frac{1}{\sqrt{M_{c}}}\sum_{j=1}^{M_{c}}\ket{c}\ket{j}\ket{\| x_{c,j} - \mu_c \|}\ket{x_{c,j} - \mu_c }$. As in the initialization procedure of Section III, we use the methods of \cite{Lloyd13,Grover,Kaye,Soklakov} to obtain the state $%
\ket{\chi_c}=\frac{1}{\sqrt{A_{c}}}\sum_{j=1}^{M}\sum_{j=1}^{M_{c}}\| x_{c,j} - \mu_c \|\ket{c}\ket{j}\ket{\| x_{c,j} - \mu_c \|}\ket{ x_{c,j} - \mu_c }
$. Tracing over $\ket{j}$ in the outer product $\ket{\chi_c}\bra{\chi_c}$
then yields the operator $\Sigma_{c}$ in time $O(\log (MN))$.

Given an input vector $\ket{x}$ in quantum form, we now present a method to
find the maximum among the $k$ discriminant functions $\delta_c(x)$ given in Eq.
(8). First, we apply $\Sigma_c^{-1}$ to the class mean $\ket{\mu_c}$, using
the matrix inversion algorithm of \cite{HHL}. As before, to avoid
exponential complexity with small eigenvalues, we introduce the pre-defined
effective condition number $\kappa_{\text{eff}}$ to limit the range of
considered eigenvalues. By \cite{HHL}, we can thus construct the state

\begin{equation}
\ket{\Sigma_c^{-1}\mu_c}
\end{equation}

\noindent for each class $c$ in time $O(\log(MN) \kappa_{\text{eff}}^3 /
\epsilon^3)$. Next, recognizing the first two terms of the discriminant
function of (8) as an inner product, we perform a SWAP test \cite{Nielsen} on
the states $\ket{\Sigma_c^{-1}\mu_c}$ and $\ket{x - \frac{1}{2} \mu_c}$ to
obtain the value

\begin{equation}
x^T\Sigma_c^{-1}\mu_c - \frac{1}{2} \mu_c^T \Sigma_c^{-1}\mu_c.
\end{equation}

\noindent This inner product evaluation requires time $O(\log(N))$. Finally,
for each class $c$, we add to the value in (24) the class prior $\pi_c =
M_c/M$. We hence obtain the discriminant values $\delta_c(x)$ for all of the $k$ classes in time $O(k \log(MN) \kappa_{\text{eff}}^3 / \epsilon^3)$. It is
then straightforward to identify the class yielding the maximum
discriminant, to which the input vector is then classified. Note that our
algorithm can be easily applied to perform quantum LDA classification, by using an
operator proportional to the scatter matrix $S_W$ (see Section III, initialization) in place
of $\Sigma_c$ for each class.

\vspace{5.5mm} \noindent\fbox{
\parbox{6.85in}{
\vspace{3.5mm}
\begingroup
\leftskip2.6mm
\rightskip\leftskip
{\bf \underline{Algorithm 2: Quantum Discriminant Analysis Classifier}}

\vspace{3mm}
\noindent
{\it Step 1}: Initialization. By querying the quantum RAM or oracles, construct the Hermitian positive semidefinite operators $\Sigma_c$ in time $O(k\log(MN))$ for all classes $c$.

\vspace{2mm}
\noindent
{\it Step 2}: Since $\Sigma_c$ is Hermitian positive semidefinite, use the method of \cite{Lloyd14} to exponentiate this operator. Apply the inversion algorithm of \cite{HHL} on the state $\ket{\mu_c}$ for each class to construct the states given by Eq. (23) in time $O(k \log (MN) \kappa_{\text{eff}}^3 / \epsilon^3)$.

\vspace{2mm}
\noindent
{\it Step 3}: Take the inner product of $\Sigma_c^{-1} \mu_c$  with $x-\frac{1}{2} \mu_c$ using the SWAP test, yielding the value in Eq. (24). This step requires time $O(\log N)$.

\vspace{2mm}
\noindent
{\it Step 4}: For each class $c$, add the class prior $\pi_c = M_c / M$ to the value in (24) to obtain the final discriminant value $\delta_c(x)$.

\vspace{2mm}
\noindent
{\it Step 5}: Select the class $c$ yielding the maximum discriminant value, and classify $x$ accordingly in time $O(k)$.

\endgroup
\vspace{3.5mm}
} }

\vspace{3mm}

\subsection{Algorithmic Complexity for Classification}

Algorithm 2 above shows the pseudocode for our quantum QDA
classifier. Step 1 (initialization) takes time $O(k\log(MN)/\epsilon)$.
Applying the inversion algorithm of \cite{HHL} for each class then takes
time $O(k \log (MN) \kappa_{\text{eff}}^3 / \epsilon^3)$. Computing all of
the inner products given by (24) requires time $O(k\log N)$, and adding the
class priors requires time $O(k)$. Finally, selecting the class with maximum
discriminant takes time $O(k)$. We add these to give the overall complexity
below:

\vspace{4mm} \noindent \textbf{Theorem 3:} \textit{The quantum QDA
classifier algorithm presented in this paper (with pseudocode given by
Algorithm 2) can be implemented in time logarithmic in both the number of
input vectors $M$ and the input vector dimension $N$. Specifically, it has a
runtime of}

\begin{equation}
O\left(k\log (MN) \kappa_{\text{eff}}^3 / \epsilon^3\right)
\end{equation}

\noindent \textit{where $k$ is the number of classes for classification, and
$\kappa_{\text{eff}}$ is a pre-defined condition number restricting the
range of eigenvalues considered for phase estimation, typically taken to be $%
O(1/\epsilon)$.}

\vspace{4mm}
Note that the complexity presented in Eq. (25) is logarithmic in both $M$
and $N$ regardless of training vector sparseness.

\section{V.  Discussion}

In this paper, we have presented a generalized algorithm from \cite{HHL} for
implementing Hermitian matrix chain operators, and applied it to implement
an algorithm for quantum LDA in polylogarithmic time. As demonstrated by
classical works such as \cite{Belhumeur, Cheng}, LDA is a powerful tool for
dimensionality reduction in fields such as machine learning and big data
analysis. Although our performance in terms of error $\epsilon $ is poorer
than classical algorithms (polynomial instead of logarithmic in $1/\epsilon $%
), we believe that this is acceptable, since it is unlikely that someone
desiring extreme levels of precision will wish to perform significant
dimensionality reduction like that provided by LDA. Rather, we believe that
the exponential speedup in terms of the parameters $M$ and $N$ should be
more significant in reducing the overall algorithmic runtime.

Our work has also presented a quantum algorithm providing exponential
speedup for the LDA and QDA classifiers. As classical studies \cite%
{Alkan,Dixon} have shown, these classifiers typically perform just as well
in terms of accuracy as the SVM (for which a quantum algorithm has been
developed, \cite{LloydSVM}). However, they tend to have much better model
stability \cite{Dixon}, which can make them more robust in face of training
data errors. Finally, discriminant analysis methods are much simpler when
generalizing to multi-class classification, whereas the SVM is more suited
for binary classification \cite{Li}. In conclusion, this work has provided
efficient exponential speedup for two important algorithms for
dimensionality reduction and classification in big data analysis.

\section{Appendix A: Proof of Theorem 1}

In this appendix, we present the derivation of Eq. (19) from Theorem 1 in
Section III. The proof of the theorem closely follows the matrix inversion
algorithm presented in \cite{HHL} (referred to in the following as the HHL
algorithm). The HHL algorithm begins with the initial state

\begin{equation}
\ket{{\psi}_0} \coloneqq \sum_{\tau = 0}^{T-1} \sin \frac{\pi(\tau + 1/2)}{T}
\ket{\tau}
\end{equation}

\noindent for large $T$ (see \cite{HHL} for more details on original
algorithm, we make a sketch below). Since the original algorithm was
designed to apply the inverse of a matrix $A$ on a specific vector $\ket{b}$%
, HHL considers the state $\ket{{\psi}_0} \otimes \ket{b}$. Here, we are
interested instead in obtaining an operator for (17), so we use the density
operator $\rho_0 = \frac{1}{\sqrt N} \sum_{i=1}^{N} \ket{i}\bra{i}$
(proportional to the identity) in place of $\ket{b}$, and we use $%
\ket{\psi_0}\bra{\psi_0}$ in place of $\ket{\psi_0}$.

Following HHL, decompose $\rho_0$ in the eigenvector basis using phase
estimation. Denote the eigenvectors of $A_1$ by $\{\ket{u_{1l}}\}_{l=1}^{N}$%
, and let $\{\lambda_{1l}\}_{l=1}^{N}$ be the corresponding eigenvalues.
Then, we write

\begin{equation}
\rho_0 = \sum_{l,l^{\prime }=1}^N \beta_{ll^{\prime }} \ket{u_{1l}} %
\bra{u_{1l'}}.
\end{equation}

Quantum phase estimation is then applied on $\rho _{0}$ for time $%
t_{0}=O(\kappa _{1}/\epsilon )$ to obtain the state

\begin{equation}
\rho _{0}^{\prime }\approx \sum_{l,l^{\prime }=1}^{N}\beta _{ll^{\prime }}%
\ket{\lambda_{1l}}\bra{\lambda_{1l'}}\ket{u_{1l}}\bra{u_{1l'}}.
\end{equation}

\noindent up to a tolerance error $\epsilon $ ($\kappa _{1}$ is the
condition number of the matrix $A_{1}$, or the ratio of the largest to
smallest eigenvalue). In this step, the exponentiation of the Hermitian
operator $A_{1}$ is performed using the trick presented in \cite{Lloyd14}.
By \cite{Lloyd14}, $n=O(\kappa _{1}^{2}/\epsilon ^{3})$ copies of $A_{1}$
are required to perform the phase estimation to error $\epsilon $, so if $%
A_{1}$ can be constructed in time $O(X)$, this step requires time $%
O(nX)=O(X\kappa _{1}^{2}/\epsilon ^{3})$.

HHL then add an ancilla and perform a unitary controlled on the eigenvalue
register. Here, we generalize this step to a controlled rotation from $%
\ket{0}\bra{0}$ to the state $\ket{\psi_{\lambda_{1l}}}\bra{\psi_{%
\lambda_{1l'}}}$, where

\begin{equation}
\ket{\psi_{\lambda_{1l}}} = \sqrt{1-C^2 f_1(\lambda_{1l})^2} \ket{0} +
Cf_1(\lambda_{1l}) \ket{1}
\end{equation}

\noindent and $C$ is a constant of order $O(\min_l
\{|f_1(\lambda_{1l})|^{-1}\})$ for normalization. Assuming $f_1$ has a
convergent Taylor series, it is possible to efficiently perform this
rotation using controlled gates (see Appendix B). This results in the
overall state

\begin{equation}
\rho_0^{\prime \prime }= \sum_{l, l^{\prime }= 1}^{N} \beta_l
\beta_{l^{\prime }} \ket{\lambda_{1l}} \bra{\lambda_{1l'}} \ket{u_{1l}} %
\bra{u_{1l'}} \ket{\psi_{\lambda_{1l}}} \bra{\psi_{\lambda_{1l'}}}
\end{equation}

Next, following HHL, we undo phase estimation to uncompute the eigenvalue
register, resulting in the state

\begin{equation}
\rho_0^{\prime \prime \prime }= \sum_{l, l^{\prime }= 1}^{N} \beta_l
\beta_{l^{\prime }} \ket{u_{1l}} \bra{u_{1l'}} \ket{\psi_{\lambda_{1l}}} %
\bra{\psi_{\lambda_{1l'}}}
\end{equation}

Finally, as in HHL, measure the ancilla to be $\ket{1}\bra{1}$. By choice of
$C$, this success probability is easily seen to be \footnote{%
As in complexity analysis, the notation $F(x) = \Omega(G(x))$ is used in
place of the expression ``$\exists$ a constant $c$ such that $F(x) \geq cG(x)$
for all x.''}

\begin{equation}
\Omega\left(\frac{\min_l |f_1(\lambda_{1l})|^2}{\max_l |f_1(\lambda_{1l})|^2}%
\right).
\end{equation}

\noindent This produces the density operator proportional to $\rho_1 =
f_1(A_1) \mathbf{I} (f_1(A_1))^\dagger$ in runtime

\begin{equation*}
O\left(\frac{X}{\epsilon^3} \cdot \kappa_1^2 \cdot \frac{\max_l
|f_1(\lambda_{1l})|^2}{\min_l |f_1(\lambda_{1l})|^2}\right).
\end{equation*}

To generalize this to the entire matrix chain, we can simply repeat this
algorithm with $A_2$ as the matrix, $f_2$ as the function, and start with
the state $\rho_1$ instead of $\rho_0$. More generally, at each iteration $j$%
, use $A_j$, $f_j$ on $\rho_{j-1}$. At each step, $n =
O(\kappa_j^2/\epsilon^3)$ copies of $A_j$ are required, and the probability
of success in measuring the ancilla is given by the expression analogous to Eq. (32). Hence, for $k$
matrices, the Hermitian operator in Eq. (17) can be implemented in time

\begin{equation}
O\left(\frac{X}{\epsilon^{3}} \cdot \sum_{j=1}^k \kappa_j^2 \cdot
\prod_{j=1}^{k} \left( \frac{\max_l |f_j(\lambda_{jl})|}{\min_l
|f_j(\lambda_{jl})|}\right)^2 \right).
\end{equation}

By \cite{HHL}, amplitude amplification can be used on the first matrix $A_1$
only to increase the measurement success probability from $\Omega\left(\frac{%
\min_l |f_j(\lambda_{jl})|^2}{\max_l |f_j(\lambda_{jl})|^2}\right)$ to $%
\Omega\left(\frac{\min_l |f_j(\lambda_{jl})|}{\max_l |f_j(\lambda_{jl})|}\right)$%
. This reduces the complexity slightly to

\begin{equation}
O\left(\frac{X}{\epsilon^{3}} \cdot \sum_{j=1}^k \kappa_j^2 \cdot \frac{%
\max_l |f_1(\lambda_{1l})|}{\min_l |f_1(\lambda_{1l})|} \cdot
\prod_{j=2}^{k} \left( \frac{\max_l |f_j(\lambda_{jl})|}{\min_l
|f_j(\lambda_{jl})|}\right)^2 \right).
\end{equation}

\noindent In the case where $A_1, ... A_k$ are density matrices presented in
a quantum RAM, $X = O(\log(N))$, and we obtain Eq. (18).

\section{Appendix B: Performing the controlled rotation of (29)}

In this section, we show how to perform a controlled rotation in the form of
(29) for an arbitrary function $f$ with convergent Taylor series.
Specifically, we are to rotate the ancilla by the angle $\theta (\lambda
)=\sin ^{-1}(Cf(\lambda ))$. This is not trivial, even for simple cases such
as $f(x)=1/x$ in the original HHL algorithm. HHL do not provide a
decomposition of this rotation in terms of controlled gates. Here, we will
present such a method for arbitrary $f$.

The idea behind our method is to approximate $\theta(\lambda)$ by its Taylor
series. We will first construct the register $\ket{f(\lambda)}$ from $%
\ket{\lambda}$, and then construct $\ket{\theta(\lambda)}$ from $%
\ket{f(\lambda)}$. This procedure is outlined below and presented in detail
in Algorithm 3.

Since $f$ has a convergent Taylor series, we approximate

\begin{equation}
f(\lambda) \approx f(x_0) + f^{\prime }(x_0)(\lambda - x_0) + ... + \frac{
f^{(n)}(x_0)(\lambda - x_0)^n}{n!}
\end{equation}

\noindent for some constant $x_0$ near $\lambda$ to order $n$. Now, by
choice of $C$, $|Cf(\lambda)| < 1$ lies in the radius of convergence of the
Maclaurin series for $\sin^{-1}(x)$. Hence, we can substitute $Cf(\lambda)$
into this Maclaurin series and approximate

\begin{equation}
\theta(\lambda) = \sin^{-1}(Cf(\lambda)) \approx Cf(\lambda) + \frac{%
(Cf(\lambda))^3}{6} + \frac{3(Cf(\lambda))^5}{40} + \frac{5(Cf(\lambda))^7}{%
112} + ...
\end{equation}

The multiplication in steps 1 and 2 on the quantum registers in Algorithm 3
can be performed very similarly to the exponentiation $a^n$ in Shor's
algorithm \cite{Shor}. Suppose the binary strings $\ket{A}$ and $\ket{B}$
are given in quantum form. Then, the classical grade-school multiplication
algorithm can be carried out by performing addition operations controlled on
the qubits representing $\ket{B}$: at the $k^{\text{th}}$ iteration, if the $%
k^{\text{th}}$ qubit of $\ket{B}$ from the right is $\ket{1}$, add $\ket{2^k
A}$ (obtained by left-shifting qubits) to the result. Repeating for all $k$
between 0 and the number of qubits in $\ket{B}$ minus one gives the desired
product.

Finally, once we have the binary representation of the rotation angle in the
register $\ket{\theta(\lambda)}$, we can implement the controlled rotation
of the ancilla. Specifically, for each term $2^r$ appearing in this binary
expansion, add a unitary controlled on the qubit coefficient of this term to
rotate the ancilla by $2^r$. Since any two-qubit controlled-$U$ operation
can be implemented with two controlled-NOT gates and single-qubit unitaries
\cite{Mermin}, the desired rotation of (29) can be implemented efficiently.

\vspace{5.5mm} \noindent\fbox{
\parbox{6.85in}{
\vspace{3.5mm}
\begingroup
\leftskip2.6mm
\rightskip\leftskip
{\bf \underline{Algorithm 3: Constructing $\ket{\theta(\lambda)}$ from $\ket{\lambda}$}}

\vspace{3mm}
\noindent
{\it Step 1}: Initialization. Prepare an auxiliary register to hold the value $\lambda-x_0$. Prepare three more working registers: the first (initialized to 1) will hold the current power of $\lambda-x_0$, the second (initialized to 1) will hold the value of the first register multiplied by the Taylor coefficient $f^{(k)}(x_0)/k!$, and the third (initialized to 0) will be a running total for the right hand side of Eq. (35).

\vspace{2mm}
{\it Step 2}: Multiply the first working register by $\lambda-x_0$ from the auxiliary.

\vspace{2mm}
\noindent
{\it Step 3}: Multiply the value in the first register by the $k^{\text{th}}$ Taylor coefficient $f^{(k)}/k!$ and store the result in the second working register.

\vspace{2mm}
\noindent
{\it Step 4}: Add the value in the second working register to the third working register.

\vspace{2mm}
\noindent
{\it Step 5}: Repeat steps 2-4 for each value $k = 1, 2, ... n$. After this step, we have successfully obtained the value $\ket{f(\lambda)}$ in the third working register.

\vspace{2mm}
\noindent
{\it Step 6}: Repeat steps 1-5 now with a register containing $\ket{Cf(\lambda)}$ in place of $\ket{\lambda}$, and with the function $\sin^{-1}$ in place of $f$. It suffices to expand around $x_0=0$. This step yields the register $\ket{\theta(\lambda)}$, by the Maclaurin series approximation of (36).

\endgroup
\vspace{3.5mm}
} }

\vspace{3mm}

{\bf Acknowledgment}: We thank the Institute of Interdisciplinary Information Sciences (IIIS) of Tsinghua University for hospitality to host our visit during which this work is done. LMD acknowledges in addition support from the IARPA MUSIQC program, the AFOSR and the ARO MURI program.


\vspace{4mm}

\begin{thebibliography}{99}
\bibitem{Welling} M. Welling, Fisher Linear Discriminant Analysis,
unpublished.

\bibitem{HHL} A.W. Harrow, A. Hassidim, and S. Lloyd,  \textit{Phys. Rev.
Lett.} \textbf{103}, 150502 (2009).

\bibitem{LloydSVM} P. Rebentrost, M. Mohseni, and S. Lloyd, \textit{%
Phys. Rev. Lett.} \textbf{113}, 130503 (2014).

\bibitem{Lloyd13} S. Lloyd, M. Mohseni, and P. Rebentrost, arXiv:quant-ph/1307.0411v2 (2013).

\bibitem{Lloyd14} S. Lloyd, M. Mohseni, and P. Rebentrost, \textit{Nature
Physics} \textbf{10}, 631-633 (2014).

\bibitem{Vasconcelos} N. Vasconcelos, PCA and LDA, unpublished.

\bibitem{Cai} D. Cai, X. He, and J. Han, Training Linear Discriminant
Analysis in Linear Time, in \textit{Proc. 2008 Int. Conf. on Data
Engineering (ICDE'08)} (2008).

\bibitem{Belhumeur} P. N. Belhumeur, P. H. Jo\~ao, and D. J. Kriegman.
\textit{IEEE Transactions on Pattern Analysis and Machine Intelligence}
19.7, 711 (1997).

\bibitem{Cheng} Y.-Q. Cheng, K. Liu, J.-Y. Yang, Y.-M. Zhuang, and N.-C. Gu,
Human face recognition method based on the statistical model of small sample
size, \textit{Intelligent Robots and Computer Vision X: Algorithms and
Techniques. International Society for Optics and Photonics} \textbf{1607},
85 (1992).

\bibitem{Giovannetti08} V. Giovannetti, S. Lloyd, and L. Maccone, Quantum
random access memory, \textit{Phys. Rev. Lett.} \textbf{100}, 160501 (2008).

\bibitem{Uetani} M. Uetani, T. Tateyama, S. Kohara, X. H. Han, Y. W. Chen,
S. Kanasaki, A. Furukawa, and X. Wei, Computer-Aided Diagnosis of Liver
Cirrhosis Based on Multiple Statistical Shape Models, in \textit{%
International Conference on Computer Information Systems and Industrial
Applications} (2015).

\bibitem{Dai} Z. Dai, C. Yan, Z. Wang, J. Wang, M. Xia, K. Li, and Y. He,
Discriminative analysis of early Alzheimer's disease using multi-modal
imaging and multi-level characterization with multi-classifier (M3), \textit{%
NeuroImage} \textbf{59}, 2187 (2012).

\bibitem{Naik} G. Naik, S. Selvan, and H. Nguyen, Single-Channel EMG
Classification With Ensemble-Empirical-Mode-Decomposition-Based ICA for
Diagnosing Neuromuscular Disorders., \textit{IEEE Trans. Neural Syst.
Rehabil. Eng.} (2015) (to be published).

\bibitem{Krusienski} D. J. Krusienski, E. W. Sellers, F. Cabestaing, S.
Bayoudh, D. J. McFarland, T. M. Vaughan, and J. R. Wolpaw, A comparison of
classification techniques for the P300 Speller, \textit{Neural Eng.} \textbf{%
3} 299 (2006).

\bibitem{Suarez-Cuenca} J. J. Su\'arez-Cuenca, W. Guo, and Q. Li,
Integration of Multiple Classifiers for Computerized Detection of Lung
Nodules in CT, \textit{Biomed. Eng. Appl. Basis Commun.} (2015).

\bibitem{Esener} I. I. Esener, S. Ergin, and T. Yuksel, A new ensemble of
features for breast cancer diagnosis, in \textit{38th International
Convention on IEEE Information and Communication Technology, Electronics and
Microelectronics (MIPRO)} (2015).

\bibitem{Alkan} A. Alkan and M. G{\"u}nay, Identification of EMG signals
using discriminant analysis and SVM classifier, \textit{Elsevier (Expert
Systems with Applications)} \textbf{39}, 44 (2012).

\bibitem{Dixon} S. J. Dixon and R. G. Brereton, Comparison of performance of
five common classifiers represented as boundary methods: Euclidean Distance
to Centroids, Linear Discriminant Analysis, Quadratic Discriminant Analysis,
Learning Vector Quantization and Support Vector Machines, as dependent on
data structure, \textit{Elsevier (Chemometrics and Intelligent Laboratory
Systems)}, \textbf{95}, 1 (2009).

\bibitem{Hastie} T. Hastie, R. Tibshirani, and J. Friedman, \textit{The
Elements of Statistical Learning} (Springer, 2009).

\bibitem{Ahuja} A. Ahuja and S. Kapoor, arXiv:quant-ph/9911082v1 (1999).

\bibitem{Mika} S. Mika, G. R{\"a}tsch, J. Weston, B. Sch{\"o}lkopf, and
K.-R. M{\"u}ller, Fisher Discriminant Analysis with Kernels (unpublished).

\bibitem{Cleve} R. Cleve, A. Ekert, C. Macchiavello, and M. Mosca. Quantum
Algorithms Revisited, arXiv:quant-ph/9708016 (1999).

\bibitem{Li} T. Li, S. Zhu, and M. Ogihara. Using discriminant analysis or
multi-class classification: an experimental investigation, \textit{Knowl
Info Syst}, \textbf{10}, 4 (2006).

\bibitem{Mermin} D. Mermin, \textit{Quantum Computer Science} (Cambridge,
2007).

\bibitem{Shor} P. Shor, Polynomial-Time Algorithms for Prime Factorization
and Discrete Logarithms on a Quantum Computer, \textit{SIAM J. Comput.},
26(5), 1484 (1997).

\bibitem{Nakahara}
M. Nakahara and T. Ohmi, \textit{Quantum Computing: From Linear Algebra to Physical Realizations} (CRC Press, 2008), Ch. 8.

\bibitem{Boyd}
S. Boyd and L. Vandenberghe, \textit{Convex Optimization} (Cambridge University Press, 2004), Ch. 5.

\bibitem{Grover}
L. Grover, T. Rudolpha, Creating superpositions that correspond to efficiently integrable probability distributions. arXiv: quant-ph/0208112.

\bibitem{Kaye}
P. Kaye, M. Mosca, in Proceedings of the International Conference on Quantum Information, Rochester, New York, 2001; arXiv: quant-ph/0407102.

\bibitem{Soklakov}
A. N. Soklakov, R. Schack, Phys. Rev. A 73, 012307 (2006).

\bibitem{Williams}
V. V. Williams, Multiplying matrices faster than Coppersmith-Winograd, in \textit{Proceedings of the forty-fourth ACM symposium on Theory of computing} (2012).

\bibitem{Nielsen}
M. A. Nielsen and I. L. Chuang, \textit{Quantum Computation and Quantum Information} (Cambridge University Press, 2011).

\end{thebibliography}
\end{document}